\documentclass[twocolumn,showpacs,preprintnumbers,amsmath,amssymb,prl]{revtex4}

\usepackage{graphics}
\usepackage{epsfig}
\usepackage{bm}

\def\etal{{\em et al.}}

\def\beq{\begin{equation}}
\def\eeq{\end{equation}}

\def\reff#1{(\ref{#1})}

\def\subsc#1{{\mbox{\rm\scriptsize #1}}}

\def\Wcmcm{\mbox{\rm Wcm$^{-2}$}}

\def\Tpulse{T_\subsc{p}}

\def\N3d{N_\subsc{3D}}

\def\vekt#1{\bm{#1}}

\def\vektr{\vekt{r}}

\def\ztilde{\tilde{z}}

\def\vekte{\vekt{e}}
\def\vektE{\vekt{E}}

\def\vektA{\vekt{A}}

\def\vektp{\vekt{p}}
\def\vektv{\vekt{v}}

\def\halb{\frac{1}{2}}

\def\Edach{\hat{E}}

\def\Ehat{\Edach}

\def\energy{{\cal{E}}}
\def\energykin{{\cal{E}}_\subsc{kin}}

\def\bra#1{\langle #1 \vert}
\def\ket#1{| #1 \rangle}
\def\braket#1#2{\langle #1 | #2 \rangle}

\def\imagi{\mbox{\rm i}}
\def\diff{\,\mbox{\rm d}}

\begin{document}

\title{Emergence of Classical Orbits in Few-Cycle Above-Threshold Ionization}
\date{\today}
\author{D.~Bauer}
\affiliation{Max-Planck-Institut f\"ur Kernphysik, Postfach 103980, 69029 Heidelberg, Germany}
\date{\today}

\begin{abstract}
The time-dependent Schr\"odinger equation for atomic hydrogen in few-cycle laser pulses is solved numerically. Introducing a positive definite quantum distribution function in energy-position space, a straightforward comparison of the numerical {\em ab initio} results with classical orbit theory is facilitated. Integration over position space yields directly the photoelectron spectra so that the various pathways contributing to a certain energy in the photoelectron spectra can be established in an unprecedented direct and transparent way. 
\end{abstract}


\maketitle

Tunneling ionization in strong laser fields is a prime example for nonperturbatively driven quantum systems. The complex structure in the photoelectron spectra can be interpreted in terms of interfering quantum orbits in the spirit of Feynman's path integral approach \cite{salieres,becker}. The recently achieved generation of phase-stabilized few-cycle laser pulses \cite{phstab} offers the opportunity to control  the continuum quantum dynamics of the released electrons or, vice verse, to use the electron spectra for measuring the electromagnetic field of the laser with sub-cycle time-resolution \cite{fewcycexp}. 

The hierarchy of theoretical approaches to few-cycle above-threshold ionization (FCATI) ranges from ``simple man's theory'' (SMT), considering only the classical orbits of the released electrons, via more quantitative, semi-analytical theories such as the ``strong field approximation'' (SFA) including rescattering of the electron at its parent atom and quantum orbit theory (QOT)  (see, e.g., \cite{becker} for a review), to the exact, numerical {\em ab initio} solution of the time-dependent Schr\"odinger equation (TDSE). The latter yields all observables that can possibly be measured in an experiment. However, a real, intuitive understanding of the underlying physical mechanisms can only be obtained with the help of simple approaches such as SMT. 

This work aims at providing a connection between the exact, numerical result and simple physical pictures. To that end a quantum distribution function is introduced in whose positive definite probability density wave packets oriented along classical trajectories emerge as the tunneling ionization regime is approached. The method is applied to FCATI where the dominating classical orbits are extremely sensitive to the so-called carrier-envelope phase (CEP).

Let $\ket{\Psi(t)}$ be the numerically determined exact solution to the TDSE $\imagi\partial_t  \ket{\Psi(t)} = \hat{H}(t) \ket{\Psi(t)}$ describing the outermost electron interacting with a laser field, i.e.,  
\beq H(t)=\halb[\hat{\vektp}+\vektA(t)]^2 + V(r) \label{hamiltonian} \eeq
(atomic units are used throughout). Here, $\vektA(t)$ is the vector potential of the laser field in dipole approximation (which is well justified for the laser parameters to be discussed in this work), and $V(r)$ is the atomic potential. 

The electric field of the laser is given by $\vektE(t)=-\partial_t\vektA(t)$. We assume that it is linearly polarized and has the form
\beq \vektE(t)=\Ehat(t) \vekte_z \cos(\omega t+ \phi)  \label{pulse}\eeq
with $\Ehat(t)$ being the pulse envelope covering $N$ laser cycles of period $T=2\pi/\omega$, $ \Ehat(t)=\Ehat\sin^2[\omega t/(2N)]$ for $ 0\leq t \leq  \Tpulse=NT$ and zero otherwise. The CEP (or ``absolute'') phase $\phi$ in \reff{pulse} is irrelevant for long pulses. Since recently, however, few-cycle laser pulses with stabilized CEP $\phi$ can be generated \cite{phstab,fewcycexp}. Under such conditions the phase $\phi$ strongly influences the dynamics of the released electrons \cite{phaseinfluence}.

The method introduced in this work is based on the projection technique proposed in \cite{schafer}. The component $\ket{\Phi_\gamma(\energy)}$ of the final wave function $\ket{\Psi_f}=\ket{\Psi(\Tpulse)}$ that contributes to energies within the  bin of width $2\gamma$ centered at $\energy$ is calculated as
\beq  \ket{\Phi_\gamma(\energy)}  = \hat{W}_\gamma(\energy)
\ket{\Psi_f} \label{projection}\eeq
where $\hat{W}_\gamma(\energy)={\gamma^{2^n}}/[{(\hat{H}_0-\energy)^{2^n} + \gamma^{2^n}}]$,
and $\hat{H}_0$ is the Hamiltonian without laser field. 
With increasing order $n=1,2,3,\ldots$ the energy-window $\hat{W}_\gamma(\energy)$ becomes
more and more rectangular. Numerically, the energy component
$\ket{\Phi_\gamma(\energy)}$ is calculated by solving the equation $\hat{W}_\gamma^{-1}(\energy) \ket{\Phi_\gamma(\energy)} = \ket{\Psi_f}$,
making use of
the factorization
$ (\hat{H}_0-\energy)^{2^n} + \gamma^{2^n}= \prod_{k=1}^{2^{n-1}} [\hat{H}_0-\energy +
\exp(\imagi \nu_{n,k})\gamma][\hat{H}_0-\energy -\exp(\imagi
\nu_{n,k})\gamma]$.
The phases $\nu_{n,k}$ can be easily calculated up to the desired
order $n$. The lowest order values read $\nu_{1,1}=\pi/2$, $\nu_{2,1}=3\pi/4$, $\nu_{2,2}=\pi/4$.

The probability $P_\gamma(\energy)$ to find the electron in a final state within an energy bin of width $2\gamma$, order $n$, and centered around $\energy$ is
$P_\gamma(\energy) = \bra{\Psi_f} \hat{W}_\gamma^2(\energy) \ket{\Psi_f} = \braket{\Phi_\gamma(\energy)}{\Phi_\gamma(\energy)} = \sum_m \vert \braket{\Psi_f}{m} \vert^2 \ f(\gamma,n)$
where $\ket{m}$ and $\energy_m$ are energy eigenvectors and eigenenergies of $\hat{H}_0$, respectively, and $f(\gamma,n)=\left\{{\gamma^{2^n}}/{[(\energy_m-\energy)^{2^n}+\gamma^{2^n}}]\right\}^2$ is the shape-function due to the finite energy-window. 

In our TDSE-solver the wave function is expanded in spherical harmonics $Y_\ell^m(\Omega)$. In the present work we restrict ourselves to linearly polarized laser light in dipole approximation so that only $m=m_0$ with $m_0$ the magnetic quantum number of the initial state contributes, i.e.,  
$ \Psi_f(\vektr)=\sum_\ell {R_\ell(r)} Y_\ell^{m_0}(\Omega)/r $.
Equation \reff{projection} thus yields $\ket{\Phi_\gamma(\energy)}$ in the form
$ \Phi_\gamma(\energy,\vektr)=\sum_\ell {R^{(\Phi_\gamma)}_\ell(r)} Y_\ell^{m_0}(\Omega)/r $, and $ P_\gamma(\energy) =\int\!\!\!\diff\Omega\int\!\!\!\diff r\,\, P_\gamma(\energy,\Omega,r) \label{energprob}$
with  \beq P_\gamma(\energy,\Omega,r) = \sum_{\ell\ell'} {Y_{\ell'}^{m_0}}^*(\Omega){Y_{\ell}^{m_0}}(\Omega) {R^{(\Phi_\gamma)}_{\ell'}}^*(r) R^{(\Phi_\gamma)}_\ell(r) \label{Pdistr} \eeq
the energy-resolved probability density. $P_\gamma(\energy,\Omega,r)$ may be viewed as a quantum distribution function in energy and position space. Integration over position space yields indeed the electron energy spectrum. However, integration over energy yields only approximately the spatial probability density. For our purposes, the great advantage of \reff{Pdistr} in comparison to, e.g., the Wigner distribution function, is (i) that $ P_\gamma(\energy,\Omega,r)$ is positive definite and (ii) that the energy-width $\gamma$ serves as an additional parameter, which turns out to play a crucial role for the emergence of classical orbits, as will be demonstrated in the following.

\begin{figure}
\includegraphics[width=.475\textwidth]{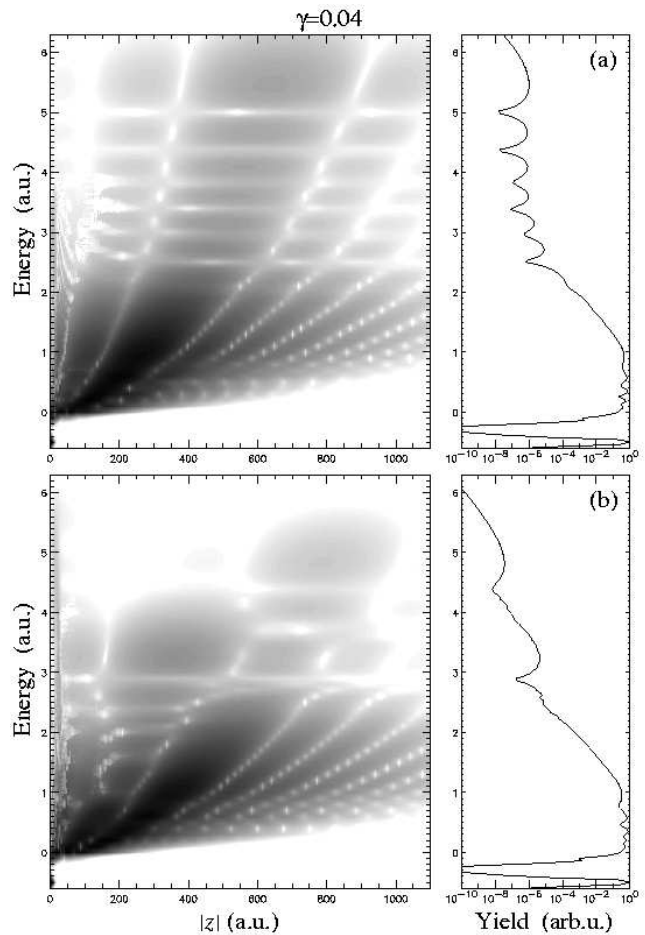}
\caption{\label{exampleI} Logarithmically scaled contour plots of $P_\gamma$-distributions ($\gamma=0.04$, $n=3$) after a 4-cycle pulse with $\phi=0$, $\omega=0.056$, $\Edach=0.0834$ in (a) $\vekte_z$-direction, i.e., $\Omega=(\vartheta,\varphi)=(0,0)$ ($\varphi$ is arbitrary for linear polarization along $\vekte_z$), and in (b) $-\vekte_z$-direction $\Omega=(\pi,0)$. Directional photoelectron spectra at the right-hand-side are obtained by integrating $P_\gamma$ over $r$ with $\vartheta$ fixed. }
\end{figure}

Figure \ref{exampleI} shows the distribution $P_\gamma(\energy,\Omega,r)$ with $\gamma=0.04$ for the electron of H(1s) after a 4-cycle pulse with CEP $\phi=0$, $\omega=0.056$ (800\,nm), and $\Edach=0.0834$ (corresponding to $2.4\times 10^{14}\,\Wcmcm$) in polarization direction $\vekte_z$ (a) and $-\vekte_z$ (b). Integration over the position coordinate yields the photoelectron spectra shown at the right-hand-side. Note that bound state energies $\energy<0$ are properly treated; the ground state population at $\energy=-0.5$ is clearly visible. Despite the short pulse duration, the spectra still show the usual pattern: after a first plateau up to $\approx 2 U_p=1.1$ (with $U_p=\Edach^2/(4\omega^2)$ the ponderomotive energy), the photoelectron yield drops down to the second plateau, reaching up to $\approx 10 U_p=5.5$. While for the electrons leaving in $\vekte_z$ direction (a), the latter plateau is well developed, the yield continues to decrease more rapidly for the electrons in the opposite direction (b). In both directions a clear interference structure is visible for energies in the second plateau. These interferences translate to the $P_\gamma$-distribution where they appear as horizontal, white stripes. In general, the energy-resolved probability density shown in Fig.\,\ref{exampleI} is rather delocalized in position space (note that more than 1000\,a.u.\ in $z$-direction are covered). This is expected since the photoelectrons contributing to a certain energy may be well described by plane waves once the laser pulse is over and they are sufficiently far away from the origin. These plane waves may interfere destructively, giving rise to the complex pattern visible in the photoelectron spectra.  The rather sharp interference patterns in the spectra of Fig.\,\ref{exampleI} are thus always accompanied by a spatially delocalized density $P_\gamma$. SMT, on the other hand, deals with classical trajectories of point-like electrons. 

Let us briefly review SMT: an electron is ``born'' at time $t_0$ with zero initial velocity and from then on moves in the laser field as if there were no Coulomb potential. In the case of a laser field polarized in $z$-direction, the velocity at times $t\geq t_0$ is $ v_z^{(0)}(t)=-\int_{t_0}^t\!\!\!\diff t'\, E(t') = A(t)-A(t_0) $
so that the final photoelectron energy and the position are given by
\beq \energykin^{(0)} = \halb {[A(\Tpulse)-A(t_0)]^2} \label{finalenerg}\eeq
and 
\beq 
z^{(0)}(t) = \ztilde(t) - \ztilde(t_0) - A(t_0)(t-t_0)  \label{finalpos}  \eeq
with $ \ztilde(t)=\int_{0}^{t} \!\!\!\diff t'\, A(t')$, respectively.

The upper index $(0)$ in \reff{finalenerg} and \reff{finalpos} indicates the ``direct'' electrons that do not interact anymore with the ion after their emission. If, instead, the electron is driven back to the origin, it may rescatter at time $t_r$ if the condition $\vert z^{(0)}(t_r)\vert \leq d_r$ is fulfilled. The smaller the distance $d_r$ is allowed to be, the less rescattering solutions exist in SMT.  
Upon rescattering electrons may assume a lateral velocity. Let $\chi$ denote the scattering angle with respect to the velocity vector of the incoming electron. The velocity after the (elastic) scattering event then reads (using cylindrical coordinates $z,\rho$) $\vektv^{(1)}(t)=[v_z^{(0)}(t_r)\cos\chi+A(t)-A(t_r)]\vekte_z  + \vert v_z^{(0)}(t_r)\vert \sin\chi\,\vekte_\rho$ with $v_z^{(0)}(t_r)=A(t_r)-A(t_0)$. For the expected electron positions 
\begin{eqnarray} z^{(1)}(t)&=&\ztilde(t)-\ztilde(t_0)-A(t_0)(t_r-t_0) \label{smtzchi}\\ 
&& +[A(t_r)(\cos\chi-1)-A(t_0)\cos\chi](t-t_r), \nonumber \\
\rho^{(1)}(t) &=& [A(t_r)-A(t_0)](t-t_r)\sin\chi \label{smtrhochi}
\end{eqnarray}
follow, and the final energy is given by $\energykin^{(1)}=\vektv^{(1)}(\Tpulse)^2/2$. In the extreme case of $180^\circ$-deflection $\chi=\pi$ one has $\energykin^{(1),(\chi=\pi)} =  {[A(\Tpulse)+A(t_0)-2A(t_r)]^2}/2$.

SMT is able to predict all the possible positions $z^{(0)}(\Tpulse)$ and $z^{(1)}(\Tpulse)$, $\rho^{(1)}(\Tpulse)$ of an electron with final energy $\energy=\energykin^{(0)}=\energykin^{(1)}$. The higher is the ionization probability at time $t_0$ (and the rescattering probability at time $t_r$) the higher should the probability to actually measure an electron at the positions predicted by SMT be.  In QOT, this is taken into account on the level of an extended SFA transition amplitude with rescattering included \cite{becker}. 

\begin{figure}
\includegraphics[width=.45\textwidth]{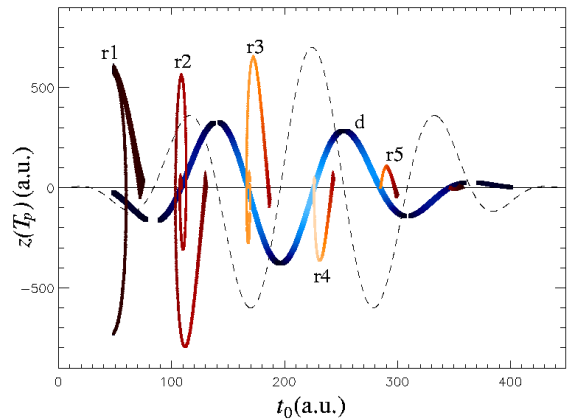}
\caption{\label{finalposfig} (Color online) Final positions $z^{(0)}(\Tpulse)$ and $z^{(1)}(\Tpulse)$ vs emission time $t_0$ for direct electrons (d) and rescattered electrons with $\chi=\pi$ (r1--r5). The shape of the laser field is indicated (dashed).  The rescattering condition was $\vert z^{(0)}(t_r)\vert \leq d_r=1$. Only emission times $t_0$ where $\vert E(t_0)\vert \geq \Ehat/10$ were taken into account in the calculation of $z^{(1)}(\Tpulse)$. The lighter the color is, the higher is the absolute value of the field amplitude at the time of emission. Same color-coding and labelling are used in Fig.\,\ref{biggergamma}.}
\end{figure}

Let us now compare the numerically determined $P_\gamma$-distribution with the SMT predictions. In Fig.\,\ref{finalposfig} the final positions $z^{(0)}(\Tpulse)$ and $z^{(1)}(\Tpulse)$ are presented as a function of the emission time $t_0$ for the laser pulse with parameters as given in the caption of Fig.\,\ref{exampleI}. The final positions $z^{(0)}(\Tpulse)$ (direct electrons) are plotted in bluish color, the final positions $z^{(1)}(\Tpulse)$ of the rescattered electrons with $\chi=\pi$ are plotted reddish. The lighter the color is, the higher was the absolute value of the electric field amplitude at the time of emission (i.e., the higher was the ionization probability). 
 
\begin{figure}
\includegraphics[width=.45\textwidth]{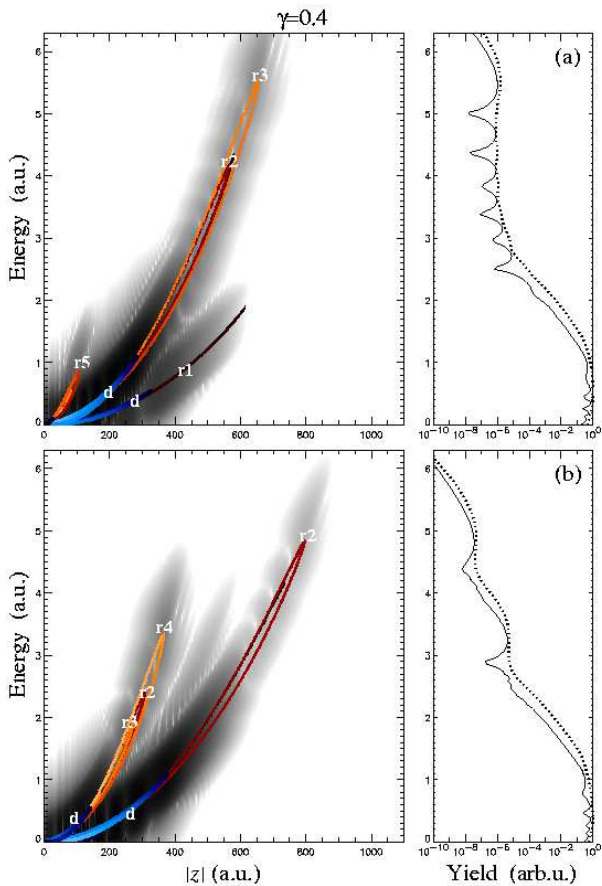}
\caption{\label{biggergamma} (Color online) Emergence of electron wave packets in the $P_\gamma$-distributions with $\gamma=0.4$, $n=3$ for the same laser pulse as in Fig.\,\ref{exampleI}. Final positions as predicted by SMT (cf.\ Fig.\,\ref{finalposfig}) are included using the same labels and color-coding. Dotted spectra at the right-hand-side were also calculated with $\gamma=0.4$.}
\end{figure}

Figure \ref{biggergamma} shows the $P_\gamma$-distributions corresponding to Fig.\,\ref{exampleI}, now calculated with a ten times wider energy-window ($\gamma=0.4$). The SMT predictions of Fig.\,\ref{finalposfig} are included using the same color coding. The larger energy width allows for rather localized electron wave packets in position space because of the uncertainty principle with respect to $\hat{H}_0$ and $\vektr$. The wave packets are clearly oriented along the classically expected positions for sufficiently high energies. The dotted spectra on the right-hand-side were calculated with $\gamma=0.4$ as well, clearly showing that the emergence of the classical orbits in the energy-resolved probability density $P_\gamma$ is accompanied by a loss of energy resolution. Compared to the solid spectra (calculated with $\gamma=0.04$), the interference pattern in the rescattering plateau is completely washed out in (a) and barely visible in (b). Instead, a constructive interference between the rescattering branches r2 and r4 is observed in the $P_\gamma$-distribution of Fig.\,\ref{biggergamma}b.

The result for the electrons in $-\vekte_z$-direction in Fig.\,\ref{biggergamma}b is particularly interesting because of the two competing rescattering solutions reaching up to high energies. As expected from SMT, the dominating solution first follows the more probable direct pathway (the lighter blue branch up to $\vert z\vert\approx 400$). At energies beyond the direct cut-off $\energy\approx 1.1$, the quantum mechanical result continues to follow this branch despite the fact that the rescattering branches r3 and r4 are more likely. This is because, contrary to the sharp cut-offs in SMT, the transition from the direct plateau in the spectrum down to the six orders of magnitude less probable rescattering regime is smooth in the quantum mechanical result. Once the level of the rescattering plateau is reached, the dominating branch is given by the SMT solution r4. The energy cut-off of r4 is again quantum mechanically extrapolated up to $\energy\approx 4.5$ before the less likely but to higher energy extending rescattering solution r2 overtakes. It is this branch r2 that defines the highest possible classical energy $\energy\approx 5$ and thus defines the ultimate classical cut-off for the electrons in $-\vekte_z$-direction. The preceding analysis nicely demonstrates the usefulness of the $P_\gamma$-distribution: through the comparison with the expected electron positions in SMT, the contribution of various pathways to each energy in the exact, numerically determined photoelectron spectrum can be understood in simple terms and great detail.

\begin{figure}
\includegraphics[width=.475\textwidth]{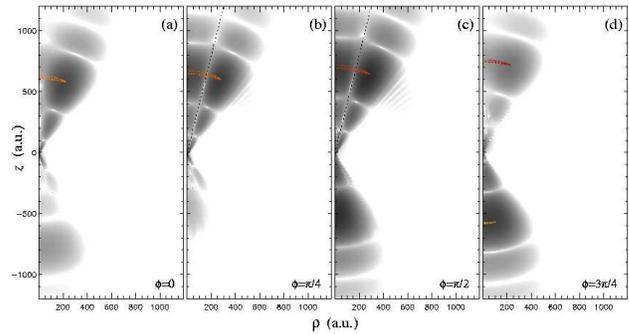}
\caption{\label{rhoz} (Color online) $P_\gamma$-distributions ($\gamma=0.04$, $n=3$) at $t=\Tpulse$ in the $\rho z$-plane for $\energy=5.0$ and (a) $\phi=0$, (b) $\pi/4$, (c) $\pi/2$, (d) $3\pi/4$. SMT solutions are included. Dotted line indicates destructive interference for $\vartheta=0.245$. }
\end{figure}

So far our study was restricted to electrons moving along the laser polarization direction. 
Figure \ref{rhoz} shows $P_\gamma$-distributions ($\gamma=0.04$) at time $t=\Tpulse$ in the $\rho z$-plane for fixed energy $\energy=5.0$ and different CEPs $\phi=0$, $\pi/4$, $\pi/2$, and $3\pi/4$. The expected SMT positions according \reff{smtzchi} and \reff{smtrhochi} are included in the plots. Note that for the first three phases $\phi$ no SMT solutions exist for electrons traveling in $-\vekte_z$-direction since $\energy=5.0$ lies beyond the corresponding classical cut-off in these cases. The classical solutions (where they exist) agree well with the quantum mechanical results. At high energies, the electron emission into narrow cones oriented along the laser polarization direction is predicted by SMT and well confirmed by the numerical result. Figure \ref{rhoz} illustrates the strong phase-dependence of few-cycle ATI. For $\phi=\pi/4$ almost no electrons of the selected energy are emitted in $-\vekte_z$ direction. For $\phi=\pi/4$ and $\pi/2$ a destructive interference for the emission angle $\vartheta=0.245$ (dotted line in (b) and (c)) is observed, which is absent for the other two phases where maximum emission of electrons with $\energy=5.0$ is {\em not} in $\vekte_z$-direction but slightly off-axis.

In summary, a positive definite quantum distribution function $P_\gamma$ with variable energy resolution $\gamma$ for the identification of relevant quantum orbits has been introduced.  Upon decreasing the energy resolution, the emergence of localized wave packets is observed if the phenomenon under investigation is accessible to (quasi) classical theory. The proposed method has been applied to the numerically determined, exact electron wave function for H(1s) in few-cycle laser pulses. Pronounced interference patterns at high energies have been observed in the photoelectron spectra. Using the $P_\gamma$-distribution, the complex structure of the electron spectra has been analyzed in terms of interfering, delocalized electrons at high energy resolution and localized electron wave packets at lower energy resolution. In that way the connection between the classical orbits of so-called ``simple man's theory'' and quantum orbit theory has been established, starting from the exact, numerically determined wave function.

The author thanks Dejan Milo\v{s}evi\'c and Wilhelm Becker for illuminating discussions. This work was supported by the Deutsche Forschungsgemeinschaft through a Heisenberg fellowship.


\end{document}